\documentclass[aps,prx,twocolumn,superscriptaddress]{revtex4}
\usepackage{xcolor}
\usepackage{graphics}
\usepackage{amsmath}
\usepackage{graphicx}
\usepackage{amsfonts}
\usepackage{amssymb}
\usepackage{float}
\usepackage{longtable}
\usepackage{epsfig}
\usepackage{latexsym}
\usepackage{theorem}
\usepackage{bbm}
\usepackage{bm}

\newtheorem{theorem}{Theorem}

\newtheorem{remark}[theorem]{Remark}

\begin{document}
\title{Conditional channel simulation}
\author{Stefano Pirandola}
\affiliation{Department of Computer Science, University of York,
York YO10 5GH, UK} \affiliation{Research Laboratory of
Electronics, Massachusetts Institute of Technology, Cambridge,
Massachusetts 02139, USA}
\author{Riccardo Laurenza}
\affiliation{Department of Computer Science, University of York,
York YO10 5GH, UK}
\author{Leonardo Banchi}
\affiliation{QOLS, Blackett Laboratory, Imperial College London, London SW7 2AZ, UK}

\begin{abstract}
In this work we design a specific simulation tool for quantum
channels which is based on the use of a control system. This
allows us to simulate an average quantum channel which is
expressed in terms of an ensemble of channels, even when these
channel-components are not jointly teleportation-covariant. This
design is also extended to asymptotic simulations, continuous
ensembles, and memory channels. As an application, we derive
relative-entropy-of-entanglement upper bounds for private
communication over various channels, including non-Gaussian
mixtures of bosonic lossy channels. Among other results, we also
establish the two-way quantum and private capacity of the
so-called \textquotedblleft dephrasure\textquotedblright\ channel.
\end{abstract}
\maketitle


\section{Introduction}

In quantum information theory~\cite{Hayashi,first}, the simulation of quantum
channels has a long history which dates back to 1996~\cite{BDSW} soon after
the introduction of the teleportation protocol~\cite{tele,telereview}. Indeed
the first idea of simulating a Pauli channel by teleporting over a two-qubit
mixed state was re-visited in various papers (e.g., see
Ref.~\cite{followPauli}). The most general formulation of channel simulation
based on local operation and classical communication (LOCC) has been given in
Ref.~\cite{PLOB} and allows one to simulate both discrete- and
continuous-variable channels~\cite{RMP}. This is also known as LOCC-simulation
of a quantum channel (see Ref.~\cite[Sec. 9]{TQC}\ for a detailed review on
the topic).

Similar ideas were put forward by Nielsen and Chuang~\cite{Array} in the
context of discrete-variable quantum computing. Ref.~\cite{Array} introduced
the notion of quantum programmable gate array (QPGA) where a channel
$\mathcal{E}$ is simulated by inserting its input $\rho$ and a program state
$\sigma$ into a universal unitary operation $G$ so that $\mathcal{E}%
(\rho)=\mathrm{Tr}_{\text{prog}}[G(\sigma_{\text{prog}}\otimes\rho)]$.
For an arbitrary channel $\mathcal{E}$ this is always possible as
long as the operation $G$ can be performed over arbitrarily many
ancillary systems (i.e., arbitrarily large programs). This can
also be understood in the context of port-based teleportation
(PBT)~\cite{PBT,PBT1,PBT2,PBT2bis,PBT3}, which allows for perfect
simulations in the limit of many ports. Indeed, PBT\ not only
provides a design for the QPGA but also shows that it can be based
on a teleportation-like LOCC, with various implications for
quantum and private communications~\cite{PBT3}.

The applications of channel simulations are various. One of the
most important is certainly the simplification of adaptive (i.e.,
feedback-based) quantum protocols into corresponding block (i.e.,
non feedback) versions. This is achieved by replacing the channels
with their simulations and to apply a suitable re-organization of
the adaptive operations of the protocol, in such a way to
decompose the output state into a tensor product of program states
up to a single quantum operation. This adaptive-to-block reduction
is also known as (teleportation) stretching of the
protocol~\cite{PLOB} and can be applied to both discrete- and
continuous-variable settings (see Ref.~\cite{TQC} for a review of
the various techniques of adaptive-to-block reduction). Combining
teleportation stretching with the relative entropy of entanglement
(REE)~\cite{REE1,REE2,REE3}, Ref.~\cite{PLOB} computed the
tightest single-letter upper bounds for the secret key capacity of
many quantum channels, also establishing the two-way quantum and
private capacities of several fundamental ones, including the
bosonic lossy channel.

In this work, we consider the general case of a quantum channel
which can be expressed as an ensemble of \textit{channel
components} with an arbitrary probability distribution. Our aim is
to design a LOCC simulation for the average channel in terms of
the single simulations associated with the various components. The
reason is because these components may have simple simulations
(e.g., with program states given by their Choi matrices) while the
average channel does not have a simple or known simulation
\textit{per se}. For instance the components may be Gaussian
channels, while the average channel can be highly non-Gaussian.
Furthermore, the channel components do not need to be jointly
teleportation covariant, which is the condition that would allow
for the direct simulation of the average channel via its Choi
matrix.

As we discuss below, this is possible by introducing a system which controls
the channel components and, therefore, creates a conditional form of channel
simulation. The state of this system will be part of the final program state
associated with the average channel. In this way, we can apply teleportation
stretching and write single-letter upper bound for the secret key capacity $K$
of the average channel in terms of the REE of the program states associated
with the single components.

As an application, we provide the finite-dimensional simulation of
a diagonal type of amplitude damping channel deriving an REE upper
bound for its $K$. We also establish $K$ and all the other two-way
assisted capacities of the \textquotedblleft
dephrasure\textquotedblright\ channel~\cite{Dephra} which is a
specific example of erasure pipeline, i.e., a channel followed by
the erasure channel. We then extend the conditional channel
simulation to bosonic channels, continuous ensembles, and memory
channels. In particular, we compute REE upper bounds for various
non-Gaussian bosonic channels which can be expressed as mixtures
of lossy channels.

\section{Simulation of channel mixtures}

\subsection{General scenario}

Let us consider a mixture of quantum channels $\mathcal{E}_{i}$ with
probability distribution $p_{i}$, i.e., the average quantum channel
\begin{equation}
\mathcal{E}=%
{\textstyle\sum\nolimits_{i}}
p_{i}\mathcal{E}_{i}~.\label{ens}%
\end{equation}
Note that channel ensembles have been considered a number of times in the
literature, including ensembles of degradable channels~\cite{Smith} and fading
channels~\cite{Satellite,PLOB}. It is clear that the Choi matrix~\cite{Choi}
$\rho_{\mathcal{E}}$\ of the average channel $\mathcal{E}$ is equal to the
convex combination of the individual Choi matrices $\rho_{\mathcal{E}_{i}}$,
i.e.,
\begin{equation}
\rho_{\mathcal{E}}=%
{\textstyle\sum\nolimits_{i}}
p_{i}\rho_{\mathcal{E}_{i}}~.
\end{equation}

Now assume that the channel $\mathcal{E}$ acts on Alice's input system $T$ and
the channel output is received by Bob. Also assume that we know the simulation
of each channel component $\mathcal{E}_{i}$, so that we may write~\cite{PLOB}%
\begin{equation}
\mathcal{E}_{i}(\rho_{T})=\mathcal{L}_{i}^{PT\rightarrow T}(\sigma_{P}%
^{i}\otimes\rho_{T}),\label{sim1}%
\end{equation}
for some trace-preserving quantum operation $\mathcal{L}_{i}$ and some program
state $\sigma_{P}^{i}$ of an extra system $P$ which can be further divided in
two subsystems $A$ (owned by Alice) and $B$ (owned by Bob). More precisely,
each $\mathcal{L}_{i}$ can always be chosen to be an LOCC~\cite{PLOB}, which
acts locally on Alice's systems $AT$ and Bob's system $B$.

In particular, if $\mathcal{E}_{i}$ is a teleportation
covariant~\cite{telecovariant} channel, then we know that $\mathcal{L}_{i}$ is
a teleportation protocol, where a Bell detection is applied to Alice's systems
$AT$ and a conditional correction unitary is applied to Bob's system $B$. In
this case, we also know that the program state $\sigma_{P}^{i}$ is equal to
the Choi matrix of the channel, i.e., $\rho_{\mathcal{E}_{i}}:=\mathcal{I}%
_{A}\otimes\mathcal{E}_{i}^{B}(\Phi_{AB})$, where $\mathcal{I}_{A}$ is the
identity channel over $A$ and $\Phi_{AB}:=\left\vert \Phi\right\rangle
\left\langle \Phi\right\vert $, with $\left\vert \Phi\right\rangle $ being the
$d$-dimensional Bell state $\left\vert \Phi\right\rangle =\sum_{i=1}^{d}%
|{ii}\rangle/\sqrt{d}$.

In the specific case where $\mathcal{L}_{i}=\mathcal{L}$ for any $i$, we call
an ensemble $\{\mathcal{E}_{i}\}$ jointly-simulable. For such an ensemble we
may write the joint simulation
\begin{equation}
\mathcal{E}(\rho_{T})=\mathcal{L}(\sigma_{P}\otimes\rho_{T}),~\sigma_{P}:=%
{\textstyle\sum\nolimits_{i}}
p_{i}\sigma_{P}^{i}. \label{kk}%
\end{equation}
In particular, the ensemble is called jointly teleportation-covariant if each
$\mathcal{E}_{i}$ is teleportation-covariant with exactly the same
teleportation LOCC $\mathcal{L}_{i}=\mathcal{L}$. In such a case we may write
Eq.~(\ref{kk}) where $\mathcal{L}$ is teleportation and the program state
becomes $\sigma_{P}=%
{\textstyle\sum\nolimits_{i}}
p_{i}\rho_{\mathcal{E}_{i}}$.

In general, the previous condition of joint simulability does not hold and it
is not known how to simulate the average channel $\mathcal{E}$ starting from
the single simulations $\{\sigma_{P}^{i},\mathcal{L}_{i}\}$ of the components
$\mathcal{E}_{i}$. We now show how this is possible by extending the idea to a
control-target scenario, where the simulations are conditional.

\begin{figure*}[pth]
\vspace{-4.5 cm}
\par
\begin{center}
\includegraphics[width=0.99\textwidth] {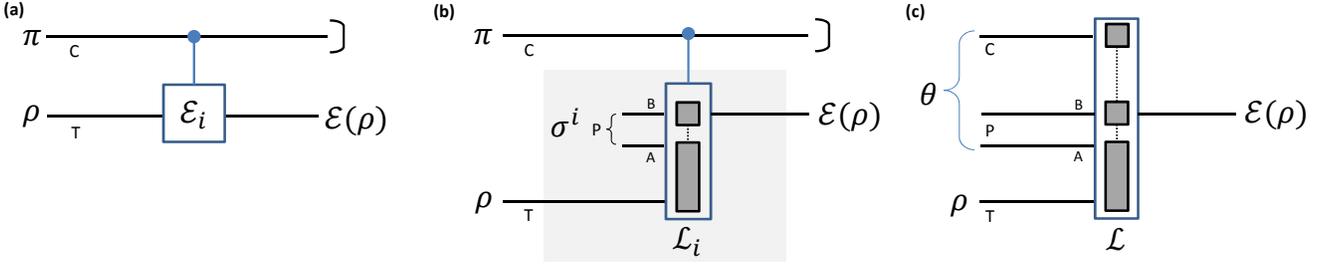}
\end{center}
\par
\vspace{-4.5cm}\caption{Steps for conditional channel simulation,
as described in Sec. II of the main text. In panels (b) and (c),
we also depict the LOCC structure of the trace-preserving quantum
operations $\mathcal{L}_{i}$ and $\mathcal{L}$.} \label{series}
\end{figure*}

\subsection{Conditional channel simulation\label{SEC_cond}}

Consider the classical state
\begin{equation}
\pi_{C}:=%
{\textstyle\sum\nolimits_{i}}
p_{i}\left\vert i\right\rangle _{C}\left\langle i\right\vert ,
\end{equation}
where $\left\vert i\right\rangle $ is the computational orthonormal basis of a
control qudit $C$ whose dimension is equal to the number $N$\ of elements in
the ensemble $\{\mathcal{E}_{i}\}$. Let us introduce the quantum operator%
\begin{equation}
M:=%
{\textstyle\sum\nolimits_{i}}
\mathcal{C}_{i}\otimes\mathcal{E}_{i},
\end{equation}
where
\begin{equation}
\mathcal{C}_{i}(\pi)=|{i}\rangle_{C}\langle i|\pi|{i}\rangle
_{C}\langle i|. \label{Cdef}%
\end{equation}
As depicted in Fig.~\ref{series}(a), we may then write
\begin{align}
\mathcal{E}(\rho_{T})  &  =\mathrm{Tr}_{C}\left[  M(\pi_{C}\otimes\rho
_{T})\right] \label{MMM}\\
&  =\mathrm{Tr}_{C}\left[
{\textstyle\sum\nolimits_{i}}
p_{i}\left\vert i\right\rangle _{C}\left\langle i\right\vert \otimes
\mathcal{E}_{i}(\rho_{T})\right]  . \label{flagged}%
\end{align}
Note that the conditional channel expression in Eq.~(\ref{flagged}) has been
also studied in the setting of degradable channels in order to show
single-letter convex decompositions of the unassisted quantum
capacity~\cite{Smith}.

Now, let us replace $\mathcal{E}_{i}$ with its simulation of Eq.~(\ref{sim1})
as also shown in Fig.~\ref{series}(b),
\begin{equation}
M(\rho_{CT})=\sum_{i}\mathcal{C}_{i}^{C\rightarrow C}\otimes\mathcal{L}%
_{i}^{PT\rightarrow T}(\sigma_{P}^{i}\otimes\rho_{CT})~.\label{tomod}%
\end{equation}
As a result, inserting the above equation into Eq.~\eqref{MMM}, we may write%
\begin{align}
\mathcal{E}(\rho_{T}) &  =\mathrm{Tr}_{C}\left[
{\textstyle\sum\nolimits_{i}}
p_{i}\left\vert i\right\rangle _{C}\left\langle i\right\vert \otimes
\mathcal{L}_{i}^{PT\rightarrow T}(\sigma_{P}^{i}\otimes\rho_{T})\right]  \\
&  =\mathcal{L}_{CPT\rightarrow T}(\theta_{CP}\otimes\rho_{T}),\label{final}%
\end{align}
where we introduce the \textquotedblleft control-program\textquotedblright%
\ state
\begin{equation}
\theta_{CP}:=%
{\textstyle\sum\nolimits_{i}}
p_{i}\left\vert i\right\rangle _{C}\left\langle i\right\vert \otimes\sigma
_{P}^{i},\label{extendedST}%
\end{equation}
and the \textquotedblleft control-program-target\textquotedblright\ LOCC
\begin{equation}
\mathcal{L}_{CPT\rightarrow T}(\rho):=\mathrm{Tr}_{C}\left[
{\textstyle\sum\nolimits_{i}}
\mathcal{C}_{i}^{C\rightarrow C}\otimes\mathcal{L}_{i}^{PT\rightarrow T}%
(\rho)\right]  ,\label{secg}%
\end{equation}
which is local with respect to systems $C$, $B$, and $AT$. The final
representation of Eq.~(\ref{final}) is also shown in Fig.~\ref{series}(c).

\subsection{Stretching and single-letter bounds\label{SEC_reas}}

We may use the channel simulation of Eq.~(\ref{final}) to stretch an adaptive
protocol of private communication over the average channel $\mathcal{E}=%
{\textstyle\sum\nolimits_{i}}
p_{i}\mathcal{E}_{i}$. Assuming that Alice and Bob have local registers
$\mathbf{a}$ and $\mathbf{b}$, and that they perform adaptive LOCCs between
each channel transmission, we may apply the teleportation stretching procedure
of Ref.~\cite{PLOB}, where each channel use is replaced with its LOCC
simulation~(\ref{final}). Considering $n$ uses of the channel, we may write
Alice and Bob's output state as
\begin{equation}
\rho_{\mathbf{ab}}^{n}=\Lambda(\theta_{CP}^{\otimes n}), \label{str22}%
\end{equation}
where $\Lambda$ is a trace-preserving LOCC including the adaptive LOCCs of the
protocol and the simulation LOCCs, while $\theta_{CP}$ is the control-program
state of Eq.~(\ref{extendedST}).

Using results from Ref.~\cite{PLOB}, we may bound the key rate achievable by
any adaptive protocol of key generation over $\mathcal{E}$. Consider an
$\varepsilon$-secure protocol with output $\rho_{\mathbf{ab}}^{n}$ where
$\left\Vert \rho_{\mathbf{ab}}^{n}-\phi_{n}\right\Vert <\varepsilon$ and
$\phi_{n}$ is a private state with $nR_{n}$ secret bits. Then, the $n$-use key
rate $R_{n}^{\varepsilon}$ must satisfy%
\begin{equation}
R_{n}^{\varepsilon}\leq\frac{E_{R}(\rho_{\mathbf{ab}}^{n})+2H_{2}%
(\varepsilon)}{(1-4\varepsilon\alpha)n},\label{rate22}%
\end{equation}
where $\alpha$ is a constant parameter associated to the dimension
$d$ of the private state $\phi_{n}$ and
$H_{2}(\varepsilon)=-\varepsilon
\log_{2}\varepsilon-(1-\varepsilon)\log_{2}(1-\varepsilon)$ is the
binary Shannon entropy. In particular, we may always choose
$\alpha$ such that $\mathrm{log}_2 d \leq \alpha n
R_{n}^{\varepsilon}$ for both discrete- and continuous-variable
systems~\cite{Chr,TQC,PLOB}. For the specific case of entanglement
distribution (so that the target state is not a private state but
a maximally entangled state of $n R_{n}^{\varepsilon}$ ebits), we
can simply set $\alpha=1$.

The previous bound is simplified by using Eq.~(\ref{str22}) and
basic properties of the REE. In fact, we may write
\begin{align}
&  E_{R}(\rho_{\mathbf{ab}}^{n})\overset{(1)}{\leq}E_{R}(\theta_{CP}^{\otimes
n})\\
&  \overset{(2)}{\leq}nE_{R}(\theta_{CP})\label{notlast}\\
&  \overset{(3)}{=}nE_{R}\left(
{\textstyle\sum\nolimits_{i}}
p_{i}\left\vert i\right\rangle _{C}\left\langle i\right\vert \otimes\sigma
_{P}^{i}\right)  \\
&  \overset{(4)}{\leq}n%
{\textstyle\sum\nolimits_{i}}
p_{i}E_{R}(\left\vert i\right\rangle _{C}\left\langle i\right\vert
\otimes\sigma_{P}^{i})\\
&  \overset{(5)}{\leq}n%
{\textstyle\sum\nolimits_{i}}
p_{i}E_{R}(\sigma_{P}^{i}),\label{last}%
\end{align}
where we have used: (1) the monotonicity of the REE under trace-preserving
LOCCs as $\Lambda$; (2) the subadditivity of the REE over tensor products; (3)
the definition of control-program state $\theta_{CP}$; (4) the convexity of
the REE over mixtures of states~\cite{Donald}; and (5) the subadditivity of
the REE over the tensor product $\left\vert i\right\rangle _{C}\left\langle
i\right\vert \otimes\sigma_{P}^{i}$ where we may always assume that the
separable state $\left\vert i\right\rangle _{C}\left\langle i\right\vert $
belongs to Alice. More precisely, let us set $P=AB$ and denote by
$\sigma_{CA|B}^{\text{sep}}$ a state which is separable with respect to the
split $CA|B$. Then, in terms of the relative entropy $S(.||.)$, we may write
\begin{align}
&  E_{R}(\left\vert i\right\rangle _{C}\left\langle i\right\vert \otimes
\sigma_{AB}^{i})\label{discard0}\\
&  =\inf_{\sigma_{CA|B}^{\text{sep}}}S(\left\vert i\right\rangle
_{C}\left\langle i\right\vert \otimes\sigma_{AB}^{i}~||~\sigma_{CA|B}%
^{\text{sep}})\\
&  \leq\inf_{\sigma_{A|B}^{\text{sep}}}S(\left\vert i\right\rangle
_{C}\left\langle i\right\vert \otimes\sigma_{AB}^{i}~||~\left\vert
i\right\rangle _{C}\left\langle i\right\vert \otimes\sigma_{A|B}^{\text{sep}%
})\\
&  =\inf_{\sigma_{A|B}^{\text{sep}}}S(\sigma_{P}^{i}~||~\sigma_{A|B}%
^{\text{sep}}):=E_{R}(\sigma_{AB}^{i}).\label{discard}%
\end{align}

By replacing Eq.~(\ref{notlast}) in Eq.~(\ref{rate22}), we therefore derive
\begin{equation}
R_{n}^{\varepsilon}\leq\frac{E_{R}(\theta_{CP})}{1-4\varepsilon\alpha}%
+\frac{2H_{2}(\varepsilon)}{(1-4\varepsilon\alpha)n}. \label{novel}%
\end{equation}
Now, by taking the limit for large $n$ and small $\varepsilon$ (weak
converse), and also using Eq.~(\ref{last}), we may write
\begin{equation}
\lim_{n,\varepsilon}R_{n}^{\varepsilon}\leq E_{R}(\theta_{CP})\leq%
{\textstyle\sum\nolimits_{i}}
p_{i}E_{R}(\sigma_{P}^{i}).
\end{equation}
Finally, by taking the supremum over all adaptive key generation protocols
$\mathcal{P}$, we get the secret key capacity of the channel%
\begin{equation}
K(\mathcal{E})=\sup_{\mathcal{P}}\lim_{n,\varepsilon}R_{n}^{\varepsilon}\leq
E_{R}(\theta_{CP})\leq%
{\textstyle\sum\nolimits_{i}}
p_{i}E_{R}(\sigma_{P}^{i}), \label{alt}%
\end{equation}
where the last inequality is expressed in terms of the REE of the
program states $\sigma_{P}^{i}$ of the channel components
$\mathcal{E}_{i}$. Recall that, for an arbitrary channel
$\mathcal{E}$, we may
write the chain of (in)equalities%
\begin{equation}
D_{2}(\mathcal{E})=Q_{2}(\mathcal{E})\leq P_{2}(\mathcal{E})=K(\mathcal{E}),
\label{capp}%
\end{equation}
where $D_{2}$ is the two-way assisted entanglement distribution capacity,
$Q_{2}$ is the two-way assisted quantum capacity, and $P_{2}$ is the two-way
assisted private capacity. Therefore, Eq.~(\ref{alt}) provides upper bounds
for all the capacities in Eq.~(\ref{capp}).

\begin{remark}
While the first inequality in Eq.~(\ref{alt}) never appeared in
the literature to our knowledge,
the final inequality $K(\mathcal{E})\leq%
{\textstyle\sum\nolimits_{i}}
p_{i}E_{R}(\sigma_{P}^{i})$ in Eq.~(\ref{alt}) can also be
obtained from a probabilistic argument where a channel component
$\mathcal{E}_{i}$\ with probability $p_{i}$ appears $np_{i}$ times
in an asymptotic protocol with large $n$. Therefore, this
component provides $np_{i}$ copies of the program state
$\sigma_{i}$\ in the stretching of the adaptive protocol. Overall,
one has the output state
$\Lambda(\otimes_{i}\sigma_{i}^{np_{i}})$~\cite{PLOB} leading to
the same final inequality as in Eq.~(\ref{alt}). \end{remark}

\begin{remark}
Because the conditional channel simulation is independent from
probabilistic/asymptotic arguments, we may write the result
directly for finite $n$. In particular, we have that
Eq.~(\ref{novel}) directly leads to the following finite-size
upper bound
\begin{equation}
R_{n}^{\varepsilon}\leq\frac{{\textstyle\sum\nolimits_{i}} p_{i}E_{R}(\sigma_{P}^{i})}{1-4\varepsilon\alpha}%
+\frac{2H_{2}(\varepsilon)}{(1-4\varepsilon\alpha)n}.
\end{equation}
for any $(n,\varepsilon,R_{n}^{\varepsilon})$-adaptive protocol
implemented over an average channel
$\mathcal{E}=\textstyle\sum\nolimits_{i} p_{i}\mathcal{E}_{i}$
with program states $\sigma_{P}^{i}$.
\end{remark}

\section{Applications in finite dimension}

\subsection{Diagonal amplitude damping channel}

Here we apply the result to a diagonal type of amplitude damping
channel (DAD) that may be represented as
\begin{equation}
\mathcal{E}_{p}^{\text{DAD}}=p\mathcal{E}_{0}+(1-p)\mathcal{E}_{1},
\label{dampdec}%
\end{equation}
where $\mathcal{E}_{0}(\rho):=\mathrm{Tr}(\rho)\left\vert
0\right\rangle \left\langle 0\right\vert $ and
$\mathcal{E}_{1}=\mathcal{I}$ is the identity channel (note that
this channel coincides with the standard amplitude damping channel
only when it is applied to the computational basis). The channel
$\mathcal{E}_{0}$ is teleportation covariant and
entanglement-breaking, so that it allows for a LOCC simulation
with a separable program state and, accordingly, $E_{R}=0$. At the
same time,
$\mathcal{E}_{1}=\mathcal{I}$ is teleportation covariant with $E_{R}%
(\rho_{\mathcal{I}})=1$. Therefore, from Eq.~(\ref{alt}), it is easy to
compute%
\begin{equation}
K(\mathcal{E}_{p}^{\text{DAD}})\leq1-p. \label{ampD}%
\end{equation}

Note that $\mathcal{E}_{0}$ are $\mathcal{E}_{1}$ are not \textit{jointly}
teleportation covariant. In fact, given a Pauli operator $P\in\{I,X,Y,Z\}$,
this is exactly commuted by the identity, but different is the case for
$\mathcal{E}_{0}$ for which we have%
\begin{equation}
\mathcal{E}_{0}(Z\rho Z^{\dagger})=Z\mathcal{E}_{0}(\rho)Z^{\dagger
},~~\mathcal{E}_{0}(X\rho X^{\dagger})=\mathcal{E}_{0}(\rho).
\end{equation}
Since the output unitaries become different for the two channel
components, these are not jointly teleportation covariant and the
DAD channel is not teleportation covariant. For this reason, we
cannot write
$K(\mathcal{E}_{p}^{\text{DAD}})\leq E_{R}(\rho_{\mathcal{E}_{p}%
^{\text{DAD}}})$. Nonetheless, since each $\mathcal{E}_{i}$ in
Eq.~\eqref{dampdec} is individually teleportation covariant, we
can use the conditional channel simulation that allows us to write
the upper bound of Eq.~\eqref{alt} in terms of the Choi matrices
of the components. The very simple form of the REE\ bound in
Eq.~(\ref{ampD}) has the advantage to make it easily extendable to
repeater chains and quantum networks~\cite{net}.

%
%

\subsection{Erasure pipeline}

Consider an arbitrary qubit channel $\mathcal{N}$ which is followed by an
erasure channel $\mathcal{E}_{p}^{\text{erase}}$ mapping the input state into
an orthogonal erasure state $|e\rangle$ with probability $p$. Explicitly we
may write the erasure pipeline $\mathcal{E}_{p}^{\text{pipe}}:=\mathcal{E}%
_{p}^{\text{erase}}\circ\mathcal{N}$ as follows%
\begin{align}
\mathcal{E}_{p}^{\text{pipe}}  &  =(1-p)\mathcal{N}+p\mathcal{E}_{e},\\
\mathcal{E}_{e}(\rho)  &  :=\mathrm{Tr}(\rho)|e\rangle\langle e|~.
\end{align}
Assume that $\mathcal{N}$ can be LOCC-simulated with a program state
$\sigma_{\mathcal{N}}$. We may write a conditional channel simulation for
$\mathcal{E}_{p}^{\text{pipe}}$ and then use Eq.~(\ref{alt}) to derive the
upper bound%
\begin{equation}
K(\mathcal{E}_{p}^{\text{pipe}})\leq(1-p)E_{R}(\sigma_{\mathcal{N}}).
\label{pipe}%
\end{equation}
Here we use the fact that the channel $\mathcal{E}_{e}$ is teleportation
covariant and entanglement-breaking ($E_{R}=0$). It is clear that
Eq.~(\ref{pipe}) also applies to a pipeline of a $d$-dimensional qudit channel
$\mathcal{N}_{d}$ followed by a $d$-dimensional erasure channel [whose output
is therefore $(d+1)$-dimensional].

\subsection{Dephrasure channel}

As an example of erasure pipeline, consider the \textquotedblleft{dephrasure
channel\textquotedblright}~\cite{Dephra}, which is a dephasing channel
$\mathcal{E}_{q}^{\text{deph}}$\ with dephasing probability $q$, followed by
an erasure channel $\mathcal{E}_{p}^{\text{erase}}$. Explicitly we may write
the dephrasure channel $\mathcal{E}_{p,q}^{\text{dr}}:=\mathcal{E}%
_{p}^{\text{erase}}\circ\mathcal{E}_{q}^{\text{deph}}$\ as follows%
\begin{equation}
\mathcal{E}_{p,q}^{\text{dr}}(\rho)=(1-p)\left[  (1-q)\rho+qZ\rho Z\right]
+p\mathcal{E}_{e}(\rho), \label{dr}%
\end{equation}
where $Z$ is the phase-flip Pauli operator. Note that the channel components
$\mathcal{E}_{q}^{\text{deph}}$ and $\mathcal{E}_{e}$ are
teleportation-covariant but not jointly. Using Eq.~(\ref{pipe}) with the fact
that the dephasing channel is simulable with its Choi matrix $\rho
_{\mathcal{E}_{q}^{\text{deph}}}$, we derive%
\begin{equation}
K(\mathcal{E}_{p}^{\text{pipe}})\leq(1-p)E_{R}\left(  \rho_{\mathcal{E}%
_{q}^{\text{deph}}}\right)  =(1-p)[1-H_{2}(q)], \label{bv1}%
\end{equation}
where $H_{2}$ is the usual binary Shannon entropy.

Now we prove that the previous relation holds with an equality. In fact,
assume that, at the output of the channel, we use a dichotomic measurement
with operators $\left\vert e\right\rangle \left\langle e\right\vert $ and
$I-\left\vert e\right\rangle \left\langle e\right\vert $. This measurement
fully decodes the second (erasure) channel $\mathcal{E}_{p}^{\text{erase}}$,
i.e., with probability $1-p$ we post-select the first (dephasing) channel
$\mathcal{E}_{q}^{\text{deph}}$. It is then known that the two-way
entanglement distribution capacity $D_{2}$ of $\mathcal{E}_{q}^{\text{deph}}$
is equal to $1-H_{2}(q)$~\cite{PLOB}. As a result, an asymptotically
achievable rate for entanglement distribution over a dephrasure channel is
equal to%
\begin{equation}
D_{2}(\mathcal{E}_{p,q}^{\text{dr}})\geq(1-p)[1-H_{2}(q)]. \label{bv2}%
\end{equation}
From Eqs.~(\ref{bv1}) and~(\ref{bv2}) we therefore conclude the exact formulas%
\begin{align}
Q_{2}(\mathcal{E}_{p,q}^{\text{dr}})  &  =D_{2}(\mathcal{E}_{p,q}^{\text{dr}%
})=P_{2}(\mathcal{E}_{p,q}^{\text{dr}})\\
&  =K(\mathcal{E}_{p,q}^{\text{dr}})=(1-p)[1-H_{2}(q)].
\end{align}

Note that we cannot achieve the lower bound in Eq.~(\ref{bv2}) using the
reverse coherent information (RCI) of the channel~\cite{RCI}. In fact, let us
write the Kraus decomposition of the dephrasure channel, which is
\begin{equation}
\mathcal{E}_{p,q}^{\text{dr}}(\rho)=\sum_{k=0}^{3}E_{k}\rho E_{k}^{\dagger},
\end{equation}
with operators
\begin{align}
E_{0}  &  =\sqrt{(1-p)(1-q)}(|0\rangle\langle0|+|1\rangle\langle1|)~,\\
E_{1}  &  =\sqrt{(1-p)q}(|0\rangle\langle0|-|1\rangle\langle1|)~,\\
E_{2}  &  =\sqrt{p}|e\rangle\langle0|,~E_{3}=\sqrt{p}|e\rangle\langle1|~.
\end{align}
We then find its Choi matrix
\begin{align}
\rho_{\mathcal{E}_{p,q}^{\text{dr}}}  &  =\frac{1-p}{2}|\Phi\rangle\langle
\Phi|+\frac{p}{2}(|0e\rangle\langle0e|+|1e\rangle\langle1e|)\\
&  -q(|00\rangle\langle11|+|11\rangle\langle00|),
\end{align}
where $|\Phi\rangle=(|00\rangle+|11\rangle)/\sqrt{2}$. As a result, we compute
the RCI of the dephrasure channel to be
\begin{equation}
I_{\text{RC}}(\mathcal{E}_{p,q}^{\text{dr}})=(1-p)[1-H_{2}(q)]-H_{2}(p)~.
\label{Defra}%
\end{equation}
This expression correctly reduces to $1-p-H_{2}(p)$ when $q=0$, which is the
RCI of the erasure channel~\cite{PLOB}. Because $\mathcal{E}_{p,q}^{\text{dr}%
}$ is not unital, we have that its RCI is different from its coherent
information, which is given by~\cite{Dephra}%
\begin{equation}
I_{\text{C}}(\mathcal{E}_{p,q}^{\text{dr}})=(1-p)[1-H_{2}(q)]-p~.
\end{equation}

\section{Extension to continuous variables}

\subsection{Asymptotic simulations}

The conditional channel simulation can be extended to ensembles of channels
having asymptotic simulations, such as bosonic channels or the amplitude
damping channel~\cite{PLOB}. This means that we may consider an average
channel $\mathcal{E}=\sum_{i}p_{i}\mathcal{E}_{i}$ where each channel
component $\mathcal{E}_{i}$ may have a generally-asymptotic LOCC simulation
as~\cite{PLOB}
\begin{equation}
\mathcal{E}_{i}(\rho_{T})=\lim_{\mu}\mathcal{L}_{i,\mu}^{PT\rightarrow
T}(\sigma_{P}^{i,\mu}\otimes\rho_{T}),\label{simCV}%
\end{equation}
where $\mathcal{L}_{i,\mu}^{PT\rightarrow T}$ is a sequence of LOCCs (between
Alice and Bob) and $\sigma_{P}^{i,\mu}$ is a sequence of program states.
Eq.~(\ref{simCV}) means that the distance between channel $\mathcal{E}_{i}$
and its asymptotic simulation, as measured by the energy-constrained diamond
norm~\cite{TQC,PLOB}, goes to zero in the asymptotic limit. For instance,
$\mathcal{E}_{i}$ may be a teleportation-covariant bosonic channel, so that we
may choose a sequence of Choi-approximating program states%
\begin{equation}
\sigma_{P}^{i,\mu}=\rho_{\mathcal{E}_{i}}^{\mu}:=\mathcal{I}\otimes
\mathcal{E}_{i}(\Phi^{\mu}),
\end{equation}
where $\Phi^{\mu}$ is a two-mode squeezed vacuum (TMSV) state with variance
$\mu$~\cite{RMP}. Perfect simulation is then obtained in the limit
$\mu\rightarrow\infty$.

In general, we may therefore write the following simulation for the average
channel%
\begin{equation}
\mathcal{E}(\rho_{T})=\lim_{\mu}\mathcal{L}_{CPT\rightarrow T}^{\mu}%
(\theta_{CP}^{\mu}\otimes\rho_{T}),\label{xx1}%
\end{equation}
where we consider a sequence of control-program states
\begin{equation}
\theta_{CP}^{\mu}:=\sum_{i}p_{i}~\left\vert i\right\rangle _{C}\left\langle
i\right\vert \otimes\sigma_{P}^{i,\mu},\label{xx2}%
\end{equation}
based on orthogonal states $\left\vert i\right\rangle $, and a sequence of
LOCCs
\begin{equation}
\mathcal{L}_{CPT\rightarrow T}^{\mu}(\rho):=\mathrm{Tr}_{C}\left[  \sum
_{i}\mathcal{C}_{i}^{C\rightarrow C}\otimes\mathcal{L}_{i,\mu}^{PT\rightarrow
T}(\rho)\right]  ,\label{xx3}%
\end{equation}
where $\mathcal{C}_{i}$ is defined in Eq.~(\ref{Cdef}).

These equations are a full extension of previous Eqs.~(\ref{final}),
(\ref{extendedST}) and~(\ref{secg}). Correspondingly, we may extend the
stretching of Eq.~(\ref{str22}) and write%
\begin{equation}
\rho_{\mathbf{ab}}^{n}=\lim_{\mu}\Lambda_{\mu}(\theta_{CP}^{\mu\otimes
n}),\label{stretchh}%
\end{equation}
for a sequence of LOCCs $\Lambda_{\mu}$~\cite{NotaBOUND}. Then, repeating the
reasonings of Sec.~\ref{SEC_reas} and using arguments from Ref.~\cite{PLOB},
we may write%
\begin{align}
E_{R}(\rho_{\mathbf{ab}}^{n}) &  =\inf_{\sigma_{\text{sep}}^{(n)}}%
S(\rho_{\mathbf{ab}}^{n}||\sigma_{\text{sep}}^{(n)})\\
&  \overset{(1)}{\leq}\inf_{\sigma_{\text{sep}}^{\mu}}S\left[  \lim_{\mu
}\Lambda_{\mu}(\theta_{CP}^{\mu\otimes n})~||~\lim_{\mu}\sigma_{\text{sep}%
}^{\mu\otimes n}\right]  \\
&  \overset{(2)}{\leq}\inf_{\sigma_{\text{sep}}^{\mu}}\underset{\mu
}{\text{\textrm{liminf}}}~S\left[  \Lambda_{\mu}(\theta_{CP}^{\mu\otimes
n})~||~\sigma_{\text{sep}}^{\mu\otimes n}\right]  \\
&  \overset{(3)}{\leq}\inf_{\sigma_{\text{sep}}^{\mu}}\underset{\mu
}{\text{\textrm{liminf}}}~S\left[  \Lambda_{\mu}(\theta_{CP}^{\mu\otimes
n})~||~\Lambda_{\mu}(\sigma_{\text{sep}}^{\mu\otimes n})\right]  \\
&  \overset{(4)}{\leq}\inf_{\sigma_{\text{sep}}^{\mu}}\underset{\mu
}{\text{\textrm{liminf}}}~S\left(  \theta_{CP}^{\mu\otimes n}~||~\sigma
_{\text{sep}}^{\mu\otimes n}\right)  \\
&  \overset{(5)}{=}n\inf_{\sigma_{\text{sep}}^{\mu}}\underset{\mu
}{\text{\textrm{liminf}}}~S\left(  \theta_{CP}^{\mu}~||~\sigma_{\text{sep}%
}^{\mu}\right)  \\
&  \overset{(6)}{\leq}n\sum_{i}p_{i}~\inf_{\sigma_{\text{sep}}^{i,\mu}%
}\underset{\mu}{\text{\textrm{liminf~}}}S\left(  \sigma_{P}^{i,\mu}%
~||~\sigma_{\text{sep}}^{i,\mu}\right)  \\
&  \overset{(7)}{=}n\sum_{i}p_{i}~E_{R}(\sigma_{P}^{i})
\end{align}
where: (1)$~\sigma_{\text{sep}}^{\mu}$ is a sequence of separable states such
that $\Vert\sigma_{\text{sep}}-\sigma_{\text{sep}}^{\mu}\Vert\overset{\mu
}{\rightarrow}0$ for separable $\sigma_{\text{sep}}$, and $\sigma_{\text{sep}%
}^{(n)}=\sigma_{\text{sep}}^{\otimes n}$ is a suboptimal choice; (2)~we use
the lower semi-continuity of the relative entropy $S$~\cite{HolevoBOOK};
(3)~we use that $\Lambda_{\mu}(\sigma_{\text{sep}}^{\mu\otimes n})$ are
specific types of separable sequences; (4)~we use the monotonicity of $S$
under $\Lambda_{\mu}$; (5) we use the additivity of $S$ over tensor products;
(6) we use the definition of $\theta_{CP}^{\mu}$ given in Eq.~(\ref{xx2}) and
the joint convexity of $S$ which can be applied by replacing $\sigma
_{\text{sep}}^{\mu}$ with $\sum_{i}p_{i}~\sigma_{\text{sep}}^{i,\mu}$ [the
orthogonal states $\left\vert i\right\rangle _{C}\left\langle i\right\vert $
can be discarded using the same arguments of Eqs.~(\ref{discard0}%
)-(\ref{discard})]; and (7) we define the REE of an asymptotic state
$\sigma:=\lim_{\mu}\sigma^{\mu}$ as follows~\cite{PLOB}
\begin{equation}
E_{R}(\sigma):=\inf_{\sigma_{\text{sep}}^{\mu}}\underset{\mu}%
{\text{\textrm{liminf}}}S\left(  \sigma^{\mu}~||~\sigma_{\text{sep}}^{\mu
}\right)  ,\label{asyREE}%
\end{equation}
with $\Vert\sigma_{\text{sep}}-\sigma_{\text{sep}}^{\mu}\Vert\overset{\mu
}{\rightarrow}0$ for separable $\sigma_{\text{sep}}$.

Using the weaker asymptotic definition of REE of Eq.~(\ref{asyREE}), we may
therefore write the upper bound
\begin{equation}
K(\mathcal{E})\leq\sum_{i}p_{i}~E_{R}(\sigma_{P}^{i}).\label{UBscop}%
\end{equation}
For computing this upper bound we need to calculate the REE of the program
states $\sigma_{P}^{i,\mu}=\sigma_{AB}^{i,\mu}$ by considering a split between
Alice ($A$) and Bob ($B$). Typically, one computes a further upper bound which
comes from picking a candidate separable state in the minimization of the REE,
i.e.,%
\begin{align}
E_{R}(\sigma_{P}^{i}) &  :=\inf_{\sigma_{\text{sep}}^{p,\mu}}\underset{\mu
}{\text{\textrm{liminf}}}S\left(  \sigma_{P}^{i,\mu}~||~\sigma_{\text{sep}%
}^{i,\mu}\right)  \label{asypp}\\
&  \leq\underset{\mu}{\text{\textrm{liminf}}}S(\sigma_{P}^{i,\mu}%
~||~\tilde{\sigma}_{\text{sep}}^{i,\mu}).\label{simplerUB}%
\end{align}
If $\sigma_{P}^{i,\mu}$ and $\tilde{\sigma}_{\text{sep}}^{i,\mu}$ are Gaussian
states, then we can use a closed formula for their relative entropy, given in
Ref.~\cite{PLOB}. Contrary to previous formulations, the formula for the
relative entropy between two arbitrary multimode Gaussian states established
in Ref.~\cite{PLOB} is directly expressed in terms of their statistical
moments, without the need of symplectic diagonalizations (for more details see
Theorem~6 and Remark 7 of Ref.~\cite{TQC}).

\subsection{Continuous ensembles}

Besides asymptotic simulations, we can also extend the tool to continuous
ensembles with associated probability densities. This means that we may
consider an average channel defined by
\begin{equation}
\mathcal{E}=\int di~p_{i}~\mathcal{E}_{i}~,
\end{equation}
where each channel component $\mathcal{E}_{i}$ may have a generally-asymptotic
LOCC simulation~\cite{PLOB}, i.e., of the form in Eq.~(\ref{simCV}). We may
extend all the previous formulas with the replacement
\begin{equation}%
{\textstyle\sum\nolimits_{i}}
p_{i}~\rightarrow\int di~p_{i}~.
\end{equation}

In particular, we may write the simulation of Eq.~(\ref{xx1}) but with a
sequence of control-program states
\begin{equation}
\theta_{CP}^{\mu}:=\int di~p_{i}~\left\vert i\right\rangle _{C}\left\langle
i\right\vert \otimes\sigma_{P}^{i,\mu},
\end{equation}
where $\left\vert i\right\rangle $ are orthogonal states, and a sequence of
LOCCs
\begin{equation}
\mathcal{L}_{CPT\rightarrow T}^{\mu}(\rho):=\mathrm{Tr}_{C}\left[  \int
di~\mathcal{C}_{i}^{C\rightarrow C}\otimes\mathcal{L}_{i,\mu}^{PT\rightarrow
T}(\rho)\right]  .
\end{equation}
This leads again to the stretching of Eq.~(\ref{stretchh}) and then to the
following upper bound%
\begin{equation}
K(\mathcal{E})\leq\int di~p_{i}~E_{R}(\sigma_{P}^{i}),
\end{equation}
where $\sigma_{P}^{i}:=\lim_{\mu}\sigma_{P}^{i,\mu}$ and $E_{R}(\sigma_{P}%
^{i})$ has the asymptotic expressions in Eqs.~(\ref{asypp})
and~(\ref{simplerUB}).

\section{Applications to non-Gaussian mixtures}

\subsection{Ensembles of lossy channels}

Let us consider the non-Gaussian average channel $\mathcal{E}:=\sum_{i}%
p_{i}\mathcal{E}_{i}$, where $\mathcal{E}_{i}:=\mathcal{E}_{\eta_{i}}$ is a
lossy channel with transmissivity $\eta_{i}$ and associated probability
$p_{i}$. The asymptotic Choi matrix of the average channel $\rho_{\mathcal{E}%
}=\lim_{\mu}\rho_{\mathcal{E}}^{\mu}$ is defined over the sequence
$\rho_{\mathcal{E}}^{\mu}=\mathcal{I}\otimes\mathcal{E}(\Phi^{\mu})$ with
$\Phi^{\mu}$ being a TMSV state. Also note that we may write%
\begin{equation}
\rho_{\mathcal{E}}^{\mu}=\sum_{i}p_{i}\rho_{\mathcal{E}_{i}}^{\mu},
\label{Choiiii}%
\end{equation}
where $\rho_{\mathcal{E}_{i}}^{\mu}$ are the quasi-Choi matrices of the single
channel components $\mathcal{E}_{i}:=\mathcal{E}_{\eta_{i}}$. Each channel
component $\mathcal{E}_{i}$ is teleportation covariant and therefore simulable
by teleporting the input over its asymptotic Choi matrix~\cite{PLOB}. More
precisely, one has the asymptotic simulation in Eq.~(\ref{simCV}) where
$\mathcal{L}_{i,\mu}^{PT\rightarrow T}$ is a generalized Braunstein-Kimble
protocol$~$\cite{teleCV} and $\sigma_{P}^{i,\mu}=\rho_{\mathcal{E}_{i}}^{\mu}$.

Note that the LOCC $\mathcal{L}_{i,\mu}^{PT\rightarrow T}$ depends on the loss
parameter $\eta_{i}$ which means that the channel components $\mathcal{E}_{i}$
are not jointly teleportation-covariant. For this reason, the simulation of
the non-Gaussian mixture $\mathcal{E}$ is not via its asymptotic Choi matrix
but can be written in the conditional and asymptotic form of Eq.~(\ref{xx1})
with $\sigma_{P}^{i,\mu}=\rho_{\mathcal{E}_{i}}^{\mu}$. Using
Eqs.~(\ref{UBscop}) and~(\ref{simplerUB}), we compute the upper bound%
\begin{equation}
K(\mathcal{E})\leq%
{\textstyle\sum\nolimits_{i}}
p_{i}\underset{\mu}{~\text{\textrm{liminf}}}~S(\rho_{\mathcal{E}_{i}}^{\mu
}||\tilde{\sigma}_{\text{sep}}^{i,\mu}),
\end{equation}
for a suitable separable Gaussian state $\tilde{\sigma}_{\text{sep}}^{i,\mu}$.
From Ref.~\cite{PLOB}, we know that the inferior limit provides the PLOB bound
$-\log_{2}(1-\eta_{i})$. Therefore, one has
\begin{equation}
K(\mathcal{E})\leq-%
{\textstyle\sum\nolimits_{i}}
p_{i}\log_{2}(1-\eta_{i}).
\end{equation}

Let us now derive a lower bound by computing the RCI of the average channel
$\mathcal{E}$ in terms of the sequence $\rho_{\mathcal{E}}^{\mu}$%
\begin{align}
I_{RC}(\mathcal{E})  &  =\lim_{\mu}I(A\langle B)_{\rho_{\mathcal{E}}^{\mu}%
},\label{uno}\\
I(A\langle B)_{\rho_{\mathcal{E}}^{\mu}}  &  =S(\rho_{A}^{\mu})-S(\rho
_{\mathcal{E}}^{\mu}),
\end{align}
where $S(.)$ is the von Neumann entropy and we have set $\rho_{A}^{\mu
}=\mathrm{Tr}_{B}\rho_{\mathcal{E}}^{\mu}$. Note that for any $\rho=\sum
_{i}p_{i}\rho_{i}$ we may use the concavity properties~\cite{NC00}%
\begin{equation}%
{\textstyle\sum\nolimits_{i}}
p_{i}S(\rho_{i})\leq S\left(  \rho\right)  \leq%
{\textstyle\sum\nolimits_{i}}
p_{i}S(\rho_{i})+H(\{p_{i}\}),
\end{equation}
where $H(\{p_{i}\}):=-\sum_{i}p_{i}\log p_{i}$ is the Shannon entropy.
Therefore, from Eq.~(\ref{Choiiii}), we may write%
\begin{gather}
I(A\langle B)_{\rho_{\mathcal{E}}^{\mu}}=S(\mathrm{Tr}_{B}\rho_{\mathcal{E}%
}^{\mu})-S(\rho_{\mathcal{E}}^{\mu})\\
=S\left(
{\textstyle\sum\nolimits_{i}}
p_{i}\mathrm{Tr}_{B}\rho_{\mathcal{E}_{i}}^{\mu}\right)  -S\left(
{\textstyle\sum\nolimits_{i}}
p_{i}\rho_{\mathcal{E}_{i}}^{\mu}\right) \\
\geq%
{\textstyle\sum\nolimits_{i}}
p_{i}S(\mathrm{Tr}_{B}\rho_{\mathcal{E}_{i}}^{\mu})-%
{\textstyle\sum\nolimits_{i}}
p_{i}S(\rho_{\mathcal{E}_{i}}^{\mu})-H(\{p_{i}\})\\
=%
{\textstyle\sum\nolimits_{i}}
p_{i}I(A\langle B)_{\rho_{\mathcal{E}_{i}}^{\mu}}-H(\{p_{i}\}).
\end{gather}
Therefore, from Eq.~(\ref{uno}) we get%
\begin{gather}
I_{RC}(\mathcal{E})=\lim_{\mu}%
{\textstyle\sum\nolimits_{i}}
p_{i}I(A\langle B)_{\rho_{\mathcal{E}_{i}}^{\mu}}-H(\{p_{i}\})\\
=%
{\textstyle\sum\nolimits_{i}}
p_{i}I_{RC}(\mathcal{E}_{i})-H(\{p_{i}\})\\
=-%
{\textstyle\sum\nolimits_{i}}
p_{i}\log_{2}(1-\eta_{i})-H(\{p_{i}\}),
\end{gather}
where we have used the fact that the RCI of the lossy channel $\mathcal{E}%
_{i}:=\mathcal{E}_{\eta_{i}}$ is simply $I_{RC}(\mathcal{E}_{i})=-\log
_{2}(1-\eta_{i})$~\cite{PLOB}. As a result, we may write the sandwich
\begin{gather}
-%
{\textstyle\sum\nolimits_{i}}
p_{i}\log_{2}(1-\eta_{i})-H(\{p_{i}\})\leq Q_{2}(\mathcal{E})\\
\leq K(\mathcal{E})\leq-%
{\textstyle\sum\nolimits_{i}}
p_{i}\log_{2}(1-\eta_{i}). \label{toGEN}%
\end{gather}

\subsection{Continuous ensembles of lossy channels}

Note that we may also consider a continuous ensemble of lossy channels with
different transmissivities, i.e., the non-Gaussian channel
\begin{equation}
\mathcal{E}:=\int d\eta~p_{\eta}\mathcal{E}_{\eta},
\end{equation}
for some suitable probability density $p_{\eta}$. It is easy to repeat
previous steps and write the upper bound%
\begin{equation}
K(\mathcal{E})\leq-\int d\eta~p_{\eta}\log_{2}(1-\eta).
\end{equation}

Another continuous ensemble of lossy channels can be created by considering a
beam splitter operation between the system and the environment
\begin{equation}
\mathcal{\tilde{E}}_{\eta}(\rho):=\mathrm{Tr}_{E}[U_{\eta}^{\text{BS}}%
(\rho\otimes\sigma_{E})U_{\eta}^{\text{BS}\dagger}]~,
\end{equation}
where in the above definition $\eta$ is the transmissivity and $\sigma_{E}$ is
a reference state of the environment. For the bosonic lossy channel
$\sigma_{E}$ is the vacuum state, while in the thermal-loss channel
$\sigma_{E}$ is a thermal state. In general, one can write any Gaussian and
non-Gaussian state using the Glauber $P$-representation
\begin{equation}
\sigma_{E}=\int d^{2}\gamma~p_{\gamma}\left\vert \gamma\right\rangle
\left\langle \gamma\right\vert \,,
\end{equation}
where $\left\vert \gamma\right\rangle $ is a coherent state with amplitude
$\gamma$. If the state $\sigma_{E}$ is classical, then $p_{\gamma}$ is a
classical probability density, and we can easily show that the non-Gaussian
channel $\mathcal{\tilde{E}}_{\eta}(\rho)$ is represented by the average%
\begin{equation}
\mathcal{\tilde{E}}_{\eta}(\rho)=\int d^{2}\gamma~p_{\gamma}\,\mathcal{E}%
_{\eta,\gamma}(\rho)~,
\end{equation}
where $\mathcal{E}_{\eta,\gamma}$ is a displaced lossy channel%
\begin{align}
\mathcal{E}_{\eta,\gamma}(\rho)  &  =\mathrm{Tr}_{E}[U_{\eta}^{\text{BS}}%
(\rho\otimes\left\vert \gamma\right\rangle _{E}\left\langle \gamma\right\vert
)U_{\eta}^{\text{BS}\dagger}]\\
&  =D(\gamma\sqrt{1-\eta})\,\mathcal{E}_{\eta,0}(\rho)\,D^{\dagger}%
(\gamma\sqrt{1-\eta}),
\end{align}
with $D(\alpha)=\exp(\alpha a^{\dagger}-\alpha^{\ast}a)$ being the
displacement operator~\cite{RMP} in terms of the ladder operators $a$ and
$a^{\dagger}$.

Let us write the beam-splitter action
\begin{equation}
U_{\eta}^{\text{BS}}a^{\dagger}U_{\eta}^{\text{BS}\dagger}=\cos\theta
a^{\dagger}-\sin\theta a_{E}^{\dagger},
\end{equation}
where $\cos^{2}\theta=\eta$ and $a^{\dagger}$ ($a_{E}^{\dagger}$) is the
creation operator acting on the system (environment). We may show that the
non-Gaussian channel $\mathcal{\tilde{E}}_{\eta}$ is teleportation covariant.
In fact, we have
\begin{equation}
\mathcal{E}_{\eta,\gamma}[D(z)\,\rho\,D(-z)]=D(z\cos\theta)\,\mathcal{E}%
_{\eta,\gamma}\left(  \rho\right)  \,D(-z\cos\theta).
\end{equation}
Since the correction unitary $D(z\cos\theta)$ does not depend on $\gamma$, we
have that the channels $\mathcal{E}_{\eta,\gamma}$ are jointly teleportation
covariant with respect to $\gamma$. As a result, $\mathcal{\tilde{E}}_{\eta}$
is teleportation covariant and simulable with its asymptotic Choi matrix
$\rho_{\mathcal{\tilde{E}}_{\eta}}=\lim_{\mu}\rho_{\mathcal{\tilde{E}}_{\eta}%
}^{\mu}$ where $\rho_{\mathcal{\tilde{E}}_{\eta}}^{\mu}=\mathcal{I}%
\otimes\mathcal{\tilde{E}}_{\eta}(\Phi^{\mu})$. Therefore, we may write the
upper bound%
\begin{equation}
K(\mathcal{\tilde{E}}_{\eta})\leq\underset{\mu}{\text{\textrm{liminf}}}%
S(\rho_{\mathcal{\tilde{E}}_{\eta}}^{\mu}||\tilde{\sigma}_{\text{sep}}%
^{\eta,\mu}),\label{keyrate}%
\end{equation}
for some suitable separable state $\tilde{\sigma}_{\text{sep}}^{\eta,\mu}$.
Note that the quasi-Choi matrix takes the form%
\begin{gather}
\rho_{\mathcal{\tilde{E}}_{\eta}}^{\mu}=\int d^{2}\gamma~p_{\gamma}%
\,\rho_{\mathcal{E}_{\eta,\gamma}}^{\mu}\\
=\int d^{2}\gamma~p_{\gamma}~[I\otimes D(\gamma\sin\theta)]\,\rho
_{\mathcal{E}_{\eta,0}}^{\mu}[I\otimes D(-\gamma\sin\theta)].
\end{gather}
Since the relative entropy does not depend on displacements, we may write%
\begin{equation}
K(\mathcal{\tilde{E}}_{\eta})\leq\underset{\mu}{\text{\textrm{liminf}}}%
S(\rho_{\mathcal{E}_{\eta,0}}^{\mu}||\tilde{\sigma}_{\text{sep}}^{\eta,\mu
})=-\log(1-\eta),
\end{equation}
so that the PLOB bound applies to the non-Gaussian channel $\mathcal{\tilde
{E}}_{\eta}$ for any classical state $\sigma_{E}$ of the environment.

\section{Extension to memory channels}

The conditional channel simulation can also be used to represent memory
quantum channels. Let us consider $M$ channel ensembles simultaneously acting
on $M$ quantum systems, i.e.,
\begin{equation}
\mathcal{E}_{\mathbf{i}}=\mathcal{E}_{i_{1}}^{1}\otimes\mathcal{E}_{i_{2}}%
^{2}\otimes\cdots\otimes\mathcal{E}_{i_{M}}^{M},
\end{equation}
where the instance $\mathbf{i}=i_{1},i_{2},\cdots,i_{M}$ occurs with joint
probability $p_{\mathbf{i}}$. The process is memoryless if and only if the
probability is factorized as $p_{\mathbf{i}}=p_{i_{1}}p_{i_{2}}\cdots
p_{i_{M}}$, otherwise there is a classical memory among the channels.

Consider the average $M$-system channel%
\begin{equation}
\mathcal{E}=%
{\textstyle\sum\nolimits_{\mathbf{i}}}
p_{\mathbf{i}}\mathcal{E}_{\mathbf{i}}~.
\end{equation}
In order to write its conditional simulation, we extend the formulas of
Sec.~\ref{SEC_cond} by means of the replacement $i\rightarrow\mathbf{i}$.
Therefore, we may write Eq.~(\ref{MMM}) where
\begin{equation}
\pi_{C}:=%
{\textstyle\sum\nolimits_{\mathbf{i}}}
p_{\mathbf{i}}\left\vert \mathbf{i}\right\rangle _{C}\left\langle
\mathbf{i}\right\vert ,~~M:=%
{\textstyle\sum\nolimits_{\mathbf{i}}}
\left\vert \mathbf{i}\right\rangle _{C}\left\langle \mathbf{i}\right\vert
\otimes\mathcal{E}_{\mathbf{i}},
\end{equation}
with $\left\vert \mathbf{i}\right\rangle =\left\vert i_{1}\right\rangle
\left\vert i_{2}\right\rangle \otimes\cdots\otimes\left\vert i_{M}%
\right\rangle $ being the computational orthonormal basis of a control system
$C$. Let us replace $\mathcal{E}_{i_{k}}^{k}$ by its simulation with program
state $\sigma_{P}^{k,i_{k}}$. Then, we may write Eq.~(\ref{final}) with the
\textquotedblleft control-program\textquotedblright\ state
\begin{equation}
\theta_{CP}:=%
{\textstyle\sum\nolimits_{\mathbf{i}}}
p_{\mathbf{i}}\left\vert \mathbf{i}\right\rangle _{C}\left\langle
\mathbf{i}\right\vert \otimes%
{\textstyle\bigotimes\nolimits_{k=1}^{M}}
\sigma_{P}^{k,i_{k}}.
\end{equation}

Assuming an adaptive protocol over $n$ uses of $\mathcal{E}$, we may write the
stretching of the output state $\rho_{\mathbf{ab}}^{n}$ as in Eq.~(\ref{str22}%
) and derive%
\begin{align}
K(\mathcal{E})  &  \leq%
{\textstyle\sum\nolimits_{\mathbf{i}}}
p_{\mathbf{i}}E_{R}\left(
{\textstyle\bigotimes\nolimits_{k=1}^{M}}
\sigma_{P}^{k,i_{k}}\right) \\
&  \leq%
{\textstyle\sum\nolimits_{\mathbf{i}}}
p_{\mathbf{i}}%
{\textstyle\sum\nolimits_{k=1}^{M}}
E_{R}\left(  \sigma_{P}^{k,i_{k}}\right)  ,
\end{align}
with suitable extensions to asymptotic simulations and continuous ensembles.

\section{Conclusions}

In this work we have designed a tool for channel simulation which
is particularly helpful for mixtures of channels. This simulation
is based on the use of a control system which generates the
probability distribution associated with the channel ensemble; the
state of this control system is then included in the final program
state. In this way we can handle mixtures of
teleportation-covariant channels which are not jointly covariant,
and we can simulate non-Gaussian channels and memory channels.

The conditional channel simulation can be exploited in the
stretching of adaptive protocols, so that we may bound the two-way
quantum and private capacities in terms of the REE. This allowed
us to establish all the two-way capacities of the recently
introduced \textquotedblleft dephrasure\textquotedblright\
channel. We have also derived bounds for various non-Gaussian
channels that can be described in terms of ensembles of lossy
channels.

Note that these bounds can also be derived by using the
probabilistic arguments of Ref.~\cite{PLOB}. However, the tool of
conditional channel simulation not only allows us to derive these
asymptotic results without probabilistic arguments but also allows
one to consider finite-size versions which are valid for any
finite number of uses.

\smallskip

\textit{Acknowledgments.}~This work have been supported by the
EPSRC via the `UK Quantum Communications Hub' (EP/M013472/1) and
by the European Union via Continuous Variable Quantum
Communications (CiViQ, Project ID: 820466). LB has been supported
by the EPSRC grant EP/K034480/1.


\begin{thebibliography}{99}
\bibitem {Choi}Recall that the Choi matrix of a quantum channel $\mathcal{E}$
is defined by propagating part of a maximally entangled state through the
channel. For a qubit channel, we therefore have $\rho_{\mathcal{E}%
}:=\mathcal{I}\otimes\mathcal{E}(\Phi)$ where $\mathcal{I}$ is the identity
channel and $\Phi:=\left\vert \Phi\right\rangle \left\langle \Phi\right\vert
$, with $\left\vert \Phi\right\rangle $ being a Bell state.
\bibitem {RCI}Recall that the RCI $I_{\text{RC}}(\mathcal{E})$ of a quantum
channel $\mathcal{E}$ is defined as the reverse coherent information when the
input state is a maximally entangled state, so that it is computed over the
channel's Choi matrix~\cite{PLOB}.
\bibitem {NotaBOUND}More precisely the limit in Eq.~(\ref{stretchh}) can be
written in trace norm if we suitably bound the energy of Alice's and Bob's
registers, a condition which can be assumed in the background of the
calculations and then relaxed at the very end. See Ref.~\cite{TQC}\ for
technical details.
\end{thebibliography}
\end{document}